\documentclass[aps,showpacs]{revtex4}
\newcommand{\Be}{\begin{equation}}
\newcommand{\Ee}{\end{equation}}
\newcommand{\Bea}{\begin{eqnarray}}
\newcommand{\Eea}{\end{eqnarray}}
\newcommand{\Ra}{\rightarrow}
\begin{document}
\setcounter{page}{0}
\begin{flushright}
{LA-UR-96-1349rev}
\end{flushright}
\vspace*{0.1in}
\begin{center}
\noindent {\bf NEUTRINO OSCILLATIONS and ENERGY-MOMENTUM CONSERVATION}
\end{center}
\thispagestyle{empty}
\mbox{}
\begin{center}
T. Goldman \\
\mbox{}
\noindent Theoretical Division \\
Los Alamos National Laboratory \\
Los Alamos, NM 87545, USA
\mbox{}\\
\vspace{2.0cm}
\mbox{}
{\bf ABSTRACT}\\
\end{center}
A description of neutrino oscillation phenomena is presented which is
based on relativistic quantum mechanics with 4-momentum conservation. 
This is different from both conventional approaches which arbitrarily use either 
equal energies or equal momenta for the different neutrino mass eigenstates. 
Both entangled state and source dependence aspects are also included.  
The time dependence of the wavefunction is found to be crucial to recovering 
the conventional result to second order in the neutrino masses. An ambiguity 
appears at fourth order which generally leads to source dependence, but the 
standard formula can be promoted to this order by a plausible convention.\\
\mbox{}\\
\mbox{}\\
PACS numbers: 14.60.Pq, 14.60.Lm, 13.10.+q, 12.15.-y\\
\newpage

\setcounter{page}{1}

\vspace{0.5cm}

{\noindent {\large {\bf I. Introduction}}}
\vspace{0.5cm}

Lipkin\cite{LIP} has raised (and continues to raise) questions regarding 
the standard derivation of neutrino oscillation phenomena\cite{PONTEC,BKnu},
which has some unsatisfactory assumptions. While it is straightforward
to verify that assuming equal energies or equal momenta both give the
same result for the oscillation length\cite{BKnu,me}, it is unclear why 
either starting point is valid.  However, it seems that the view taken in Ref.{\cite{LIP}} 
that oscillations are a momentum/spatial phenomenon and do not involve 
energy/time is at least incomplete, if not wrong:\cite{MMN} Lipkin's very first example 
claims that the energy of a $K^0$ produced in a two body reaction is fixed 
by energy conservation, so the momenta of the $K_L$ and $K_S$ components 
must be different. However, the momentum is also fixed by a conservation law. 
How then does one get around this conundrum, since the states observed in
experiments certainly propagate long enough distances to be considered
on-shell?  I present here an analysis which includes elements of covariance 
and entangled states which do not seem to have been discussed in the literature 
for the particular case of neutrino oscillations. Questions of wavepackets or 
coherence, which have been discussed in detail in Refs.(\cite{PONTEC,BKnu}), 
are applied expost facto.

\vspace{1cm}

{\noindent {\large {\bf II. Calculation }}}
\vspace{0.5cm}

We commence with a decay source of mass $M$ which emits neutrino flavor
eigenstates and consider it in its rest frame, with no momentum dispersion. 
Such a state must be infinitely spread out in space, but we will return to the 
question of packeting the wavefunction later. The {\it ket} transformation is 
then
\Bea
|M, {\vec 0}, M^2 \rangle \Ra c |\sqrt{k^2 + m_1^2}, \vec{k}, m_1^2 \rangle 
\times |E_1, - \vec{k}, M_f^2 \rangle \hspace{1.5in} \nonumber \\
+ s |\sqrt{q^2 + m_2^2}, \vec{q}, m_2^2 \rangle 
\times |E_2, - \vec{q}, M_f^2 \rangle	\label{eq1}
\Eea
where $m_{1,2}$ are the masses of the neutrino mass eigenstates, $c,s$
are the cosine and sine of the mixing angle, $\theta$, between the mass
and flavor eigenstates, $M_f$ is the mass of the recoiling final state of the 
source, $\vec{k}, \vec{q}$ are the three-momenta of the two neutrino mass 
eigenstates, and $E_{1,2}$ are the energies of the recoiling source (daughter) 
final states for each neutrino mass eigenstate.

It is important to recognize that the neutrino oscillation here is
determined by a state preparation, which is in turn defined by the
recoil mass of the entangled component. A fixed invariant mass of that
component does {\em not} define which neutrino mass eigenstate
component has been produced, but allows both mass eigenstates to
appear. It is only if we measure the energy or momentum of the recoil
entangled state separately to sufficient accuracy that projection of
one particular neutrino mass eigenstate occurs. In the normal quantum
mechanical way, usually discussed in terms of two slit {\it gedanken}
experiments, when the interference occurs, it is that of this single
neutrino wave (specified by $M_f^2$) with itself.

To keep a concrete picture in mind, one may think of the example of a
source initial state consisting of a pion and the corresponding
recoiling final state of a muon.  However, the formulation is quite
general, and simply reflects the fact that sufficiently accurate
measurement of the recoiling component of the entangled states can in
principle allow a particular mass eigenstate to be determined,
eliminating any oscillations in the usual quantum mechanical manner. If
the source is more complicated, such as a $\beta$-decaying nucleus, or
a muon, we need only integrate over the allowed range of $M_f$, which
is the invariant mass constructed from the sum of the four-momenta of
the recoiling particles: (possibly excited) final state nucleus plus
electron or `other' neutrino plus electron, in the two example cases.
The argument here is focused on interference of neutrinos for the case
with a given, fixed recoiling invariant mass because that is clearly a
definable, (radiative corrections to the charged particle recoil
notwithstanding,) and experimentally common case, as in two body weak
decay of charged pions.\cite{clear}

In the normal quantum mechanical fashion, we now implicitly integrate 
over the recoiling states (over all energies and momenta but with fixed
invariant total mass), and find the evolved neutrino {\it ket} after
time $t$ is
\Be
| \nu , t  \rangle = c |k, m_1  \rangle e^{-i(\sqrt{k^2 + m_1^2}~t - k~x)} 
+ s |q, m_2  \rangle e^{-i(\sqrt{q^2 + m_2^2}~t - q~x)}  \label{eq2}
\Ee
where we have assumed the momentum and position vectors are parallel to
simplify the notation. Using the $t=0$ definition of the flavor
eigenstate to describe the amplitude for detecting it at the event $(t,
\vec{x})$, we have
\Be
\langle \nu , 0 | \nu , t  \rangle = c^2 e^{-i(E_k~t - k~x)} 
+ s^2 e^{-i(E_q~t - q~x)}  \label{eq3}
\Ee
with the obvious simplifications in notation.  Note that the event
describes the registration of the neutrino in the experimental
detection apparatus, and has, in principle, nothing to do (for quantum
waves) with the transit time from the source to the location of the
detection in the apparatus, for sufficiently broad wavepackets.
Indeed, if the latter relation could be made precise, the interference
would be destroyed just as by the sufficiently accurate detection of
energy or momentum of the recoiling source final state.\cite{BKnu})

There may be concern here that the delta-function momenta of the 
recoiling final state components eliminates the coherence of the two 
terms in Eq.(\ref{eq2}) and so obviates Eq.(\ref{eq3}). Indeed, a 
density-matrix view would suggest as much. However, there would 
be no question about this issue if the source were in a packet of finite 
energy-momentum width, such as is necessary to localize the source 
in the first place. What is being done here is to analyze the contributions 
of such a packet by each (source) spectral line within it, so as to make 
energy-momentum conservation manifest. As a result, even if the 
recoiling state were measured with sufficient accuracy to distinguish 
the two components, still no determination could be made since it 
would remain unknown from which spectral component the momenta 
originated. Very recently, questions have been raised regarding possible 
interferences from the different spectral components of the initial 
state~\cite{GSI,pc}. That question is not addressed here. 

A few kinematical relations are now useful. Representing either
momentum case by $p$ and either neutrino mass by $m$, we have
\Be
\sqrt{p^2 + M_f^2} + \sqrt{p^2 + m^2} = M  \label{eq4}
\Ee 
which can be solved for the neutrino energy to show that  
\Be
\sqrt{p^2 + m^2} = \frac{M^2 - M_f^2 + m^2}{2 M}  \label{eq5}
\Ee
Thus, we have explicitly for the neutrino energies:
\Bea
E_k &=& \sqrt{k^2 + m_1^2} \nonumber \\
&=& \frac{M}{2} - \frac{M_f^2}{2 M} + \frac{m_1^2}{2 M}  \label{eq6}
\Eea
and 
\Bea
E_q &=& \sqrt{q^2 + m_2^2} \nonumber \\
&=& \frac{M}{2} - \frac{M_f^2}{2 M} + \frac{m_2^2}{2 M}  \label{eq7}
\Eea 
We now want the differences between these energies and momenta, and it
is convenient to expand all cases about the averages of each pair,
which we do to fourth order in the neutrino masses. For the energy 
we have
\Be
E_{\nu}^{av} = \frac{M^2 - M_f^2}{2 M} + \frac{m_1^2 + m_2^2}{4 M}
\label{eq8}
\Ee 
and for the momentum, we solve Eqs.(\ref{eq6},\ref{eq7}), for $k$ and
$q$ respectively, and perform a Taylor expansion to fourth order in the
neutrino masses
\Be
k \simeq \frac{M^2 - M_f^2}{2 M} - \frac{m_1^2}{2 M} 
\frac{(M^2 + M_f^2)}{(M^2 - M_f^2)} - 
\frac{m_1^4 M^2 M_f^2}{M (M^2 - M_f^2)^3}\label{eq6a}
\Ee 
and 
\Be
q \simeq \frac{M^2 - M_f^2}{2 M} - \frac{m_2^2}{2 M} 
\frac{(M^2 + M_f^2)}{(M^2 - M_f^2)} - 
\frac{m_2^4 M^2 M_f^2}{M (M^2 - M_f^2)^3} \label{eq7a} ,
\Ee 
then average, to obtain
\Bea
p_{\nu}^{av} = \frac{M^2 - M_f^2}{2 M} - \frac{(m_1^2 + m_2^2)}{4 M}
\frac{(M^2 + M_f^2)}{(M^2 - M_f^2)} \nonumber \\
- \frac{(m_1^4 + m_2^4)}{2 M}\frac{M^2 M_f^2}{(M^2 - M_f^2)^3}
\label{eq9}.
\Eea
We next define
\Bea
\Delta E = \frac{\Delta m^2}{2 M} \label{eq10}
\Eea
and 
\Be
\Delta p = - \frac{\Delta m^2}{2 M} \frac{(M^2 + M_f^2)}{(M^2 - M_f^2)} 
- \frac{\Delta m^2 ( m_2^2 + m_1^2) M M_f^2}{(M^2 - M_f^2)^3} \label{eq11}
\Ee
where $\Delta m^2 = m_2^2 - m_1^2$, so that 
\Be
E_{q,k} = E_{\nu}^{av} \pm  \frac{\Delta E }{2} \label{eq12}
\Ee
and
\Be
p_{q,k} = p_{\nu}^{av} \pm  \frac{\Delta p }{2} \label{eq13}.
\Ee

We now factor out the average phase quantities, to rewrite
Eq.(\ref{eq3}) as
\Be
\langle \nu , 0 | \nu , t  \rangle = e^{-i\phi} \left( c^2
e^{i\frac{(\Delta E t - \Delta p x)}{2}} + s^2 e^{-i\frac{(\Delta E t -
\Delta p x)}{2}} \right)  \label{eqint}
\Ee
where 
\Bea
\phi = (E_{\nu}^{av}t - p_{\nu}^{av}x) = \left[\frac{M^2 - M_f^2}{2
M}  - \frac{m_1^2 + m_2^2}{4 M}\right]t \hspace{2.0in} \nonumber \\
- \left[\frac{M^2 - M_f^2}{2 M} - \frac{m_1^2 + m_2^2}{4 M}\frac{M^2 +
M_f^2}{M^2 - M_f^2} - \frac{m_1^4 + m_2^4}{2 M}\frac{M^2 M_f^2}{(M^2 -
M_f^2)^3}\right]x \label{avphs}.
\Eea

Squaring Eq.(\ref{eqint}) and substituting from
Eqs.(\ref{eq10},\ref{eq11}) now gives the persistence probability,
$P(t,x)$, of the initial neutrino flavor to fourth order in the
neutrino masses as
\Bea
P(t,x) \equiv | \langle \nu , 0 | \nu , t  \rangle |^2 = c^4 + s^4
\hspace{3.0in} \nonumber \\
\hspace*{-0.1in} + 2c^2s^2 \cos \Bigg\{ \frac{\Delta m^2 t}{2 M} +
\frac{\Delta m^2 x}{2 M} \left( \frac{M^2 + M_f^2}{M^2 - M_f^2} +
\frac{2 (m_1^2 + m_2^2) M^2 M_f^2}{(M^2 - M_f^2)^3} \right) \Bigg\}
\label{eqintsq} .
\Eea

Experimentally, one usually measures $x$ and implicitly infers the
value of $t$ to compare with this formula.\cite{LIP} A question
has been raised in other contexts\cite{sws} of what relation is
the correct one to use in this regard. Here we evaluate the relation by
averaging the classical velocities of the two components in a
conventional manner, assuming, as usual, that the $m_i^2$ are tiny
compared to $M^2$, $M_f^2$, and $M^2 - M_f^2$, and that this reflects
the motion of centroids of wavepackets. To zeroth order in the neutrino
masses, this sets $t = x$ (in units where $c = 1$). We will need the
result to second order in the neutrino masses, and find it by defining
the average velocity
\Bea
v_{av} &=& \frac{1}{2} \left( \frac{k}{E_k} + \frac{q}{E_q} \right) 
\nonumber \\ 
&=& 1 - \frac{(m_1^2 + m_2^2) M^2}{(M^2 - M_f^2)^2},
\Eea
where we have used Eqs.(\ref{eq6},\ref{eq7},\ref{eq6a},\ref{eq7a}) and
then set $t = t_{av}(x)$ where
\Bea
t_{av}(x) &=& x / v_{av} \nonumber \\
&=& x \left[ 1 + \frac{(m_1^2 + m_2^2) M^2}{(M^2 - M_f^2)^2} \right] \label{tbar} . 
\Eea 
We emphasize that this is only necessary to go beyond leading order,
and note that the same result may be obtained to this order from
$p_{\nu}^{av} / E_{\nu}^{av}$. 	
	
Using the usual trigonometric identities for oscillations in
Eq.(\ref{eqintsq}) and substituting for $t$ from Eq.(\ref{tbar}), 
we finally obtain
\Bea
P(t_{av}(x),x) = 1 - \sin^2(2\theta) \sin^2 \Bigg\{ \frac{\Delta m^2
x}{4 M} \left[ \frac{2 M^2}{M^2 - M_f^2} \right. \nonumber \\
\left. \times \left( 1 + \frac{(m_1^2 + m_2^2) (M^2 + M_f^2)}{2 (M^2 -
M_f^2)^2} \right) \right]  \Bigg\} \label{eqresult} .
\Eea

\vspace{1cm}

{\noindent {\large {\bf III. Comparison with Conventional Results }}}
\vspace{0.5cm}

To compare this formula with the conventional result, obtained by
assuming that either the energies or momenta of the two neutrino
components are the same, recall that\cite{PONTEC,BKnu,me}
\Be
P_{conv.}(x) = 1 - \sin^2(2\theta) \sin^2 \left[ \frac{\Delta m^2
x}{4 k} \right] \label{eqwrong}.
\Ee
where the usual relativistic approximations are used that $k \approx
E_{\nu}$ and the signal in the detector occurs at $t = x$, and which
can be derived assuming {\em either} that the two neutrino components
have the same energy or that they have the same momentum.

To complete the comparison, we simply need to recognize that, in
Eq.(\ref{eqresult}), $2M/(M^2 - M_f^2) = 1/p_{\nu}^{av} =
1/E_{\nu}^{av}$ to leading order. Hence the result to second order in
the neutrino masses is identical to that produced by the conventional
analyses. That is,
\Be
P(x,x)^{2nd~order} = P_{conv.}(x) . 
\Ee

It is also interesting to extend the interpretation of the result,
Eq.(\ref{eqresult}), to fourth order in the neutrino masses. To do
this, we need to recognize that the correction factor multiplying the
second order result,
$$
1 + \frac{(m_1^2 + m_2^2) (M^2 + M_f^2)}{2 (M^2 - M_f^2)^2} ,
$$
is precisely that needed to promote the factor of $2M/(M^2 - M_f^2)$,
interpreted as $1/p_{\nu}^{av}$, from its zeroth order value in the
neutrino masses to second order, as can be seen from Eq.(\ref{eq9}).
This would {\em not} be true if we were to interpret the factor
$2M/(M^2 - M_f^2)$ as $1/E_{\nu}^{av}$. Thus, {\em there are no fourth
order effects at all} if we define the neutrino oscillation formula as
\Be
P(t_{av}(x),x) = 1 - \sin^2(2\theta) \sin^2 \Bigg\{ \frac{\Delta m^2
x}{4 p_{\nu}^{av}} \Bigg\} .
\Ee

One could also argue that the procedure used here of averaging the
velocities of the two neutrinos to infer a transit time is
incorrect.(Perhaps this definitional ambiguity of the {\em detection 
time} is related to what Lipkin\cite{LIP} has in mind.) Certainly
the quantum mechanical interference will disappear if the value of the
transit time were, in fact, measured to sufficiently high levels of
accuracy, as one may then separate (in principle) the two velocity
components. More simply, one could ask if the averaging should not be
weighted by the relative amplitudes (or probabilities) of the two
neutrino mass eigenstates.\cite{DVA} To illustrate the level of sensitivity, we
can look at an extreme version of this question: What if we were to
take $t~=~x$ and correspondingly identify the energy/momentum factor
($E_0 = (M^2 - M_f^2)/(2M)$) that both neutrinos would have if they
were massless? This is reasonable since the neutrino energy is
generally {\em not} known accurately in experiments. We would then have
\Bea
P(x,x) = 1 - \sin^2(2\theta) \sin^2 \Bigg\{ \frac{\Delta m^2 x}{4
E_{0}} \left[ 1 + \frac{(m_1^2 + m_2^2) M_f^2}{(M^2 - M_f^2)^2}
\right]  \Bigg\} \label{srcdep}  
\Eea
where $E_{0}$ is the energy given by Eq.(\ref{eq6}) or Eq.(\ref{eq7})
at zero neutrino mass.

The result Eq.(\ref{srcdep}) is {\em source} dependent: The coefficient
of the sum of neutrino squared masses varies by almost three orders of
magnitudes from 170~GeV$^{-2}$ for muon-neutrinos from $\pi~\Ra~\mu$ to
0.21~GeV$^{-2}$ for those from $K~\Ra~\mu$ two body decay.  In view of
the experimental bounds on the neutrino masses themselves, however, it
seems unlikely that any such higher order source term dependence effect
would be experimentally measurable in the foreseeable future.  

The discussion has been limited here to the two neutrino mixing case
for simplicity. The extension to the case of three neutrino mixing is
immediate. See for example Ref.(\cite{DVAB}).

\vspace{1cm}

{\noindent {\large {\bf IV. Conclusion }}}
\vspace{0.5cm}

Starting from a detailed representation of the neutrino source, and
fixing only the invariant mass of the recoiling component of resulting
entangled state, the evolution of the neutrino amplitude consisting of
two mass eigenstates has been shown to produce, to second order in the
neutrino eigenmasses, the same oscillation relation as obtained by the
usual (but unjustified) assumption of either equal momenta or energy
for the two neutrino components. Note that the variations due to the
time between the production of the neutrino by the source and its
detection, and the distance from the source to the point of detection
must {\em both} be included to reproduce the usual result, contrary to
the assertions in Ref.(\cite{LIP}).

Dependence on the source, through the invariant mass of the final
state, does not develop even at the level of fourth order terms in the
neutrino masses, provided the standard neutrino oscillation formula is
defined in terms of the average momentum of the two (or more) neutrino
mass eigenstates. Use of the momentum that would be carried by a zero
mass neutrino will induce apparent fourth order corrections which are
certainly negligibly small for present day experiments. Nonetheless, we
note that these effects can be determined precisely when the source
final state recoil invariant mass is  itself defined precisely, as in
the case of two body decay sources (such as pions).

The calculation discussed here has been carried out for a neutrino
source at rest in the same frame as that of the neutrino detector.
However, the boost of that source relative to the detector rest frame
can only affect the energy and momentum of the observed neutrino by the
standard relativistic transformation. Therefore the results will also
be applicable to all currently achievable experimental conditions.
The extension to cases where the neutrino energies deriving from
the source are nonrelativistic, (that is, for $M^2 - M_f^2 \approx 
m^2$,) does not seem likely to be of use at present.\cite{DVAT} 

Related discussions, in the more accessible case of neutral kaon 
oscillations, may be found in Refs.\cite{Lowe,ETC}.

{\bf Acknowledgments.} I thank D.\ V.\ Ahluwalia, M.\ M.\ Nieto, 
L.\ Okun, A.\ Gal and A.\ Merle for valuable conversations. Although 
perhaps not directly applicable, I am sure that my thinking in this 
matter has been influenced by the beautifully clear analysis by 
Kayser\cite{BK} of other particle oscillation phenomena.
This work was carried out in part under the auspices of the National 
Nuclear Security Administration of the U.S. Department of Energy at 
Los Alamos National Laboratory under Contract No. DE-AC52-06NA25396.

{\it Note added.} This paper first appeared as "Source Dependence 
of Neutrino Oscillations", hep-ph/9604357v1 but is only slightly 
modified here.  Shortly after this work was completed, Refs.\cite{fldthy,ptr}
were brought to my attention. In the first, the approach is to treat
the states involved as off-shell. From my analysis, this does not seem
to be necessary. It should be noted, however, that the paper carefully
justifies the assumption, used here, that the spatial position difference
vector from source to detection point is parallel to the momentum
vector for the interfering neutrino amplitudes. In the second, pion
decay with differing neutrino energies and momenta is considered.
However, there the emphasis is placed on the time variation of the
phase, [completely opposite to the view in Ref.(\cite{LIP})], with yet
another averaging procedure (justified by an aside on the question of
coherence) used to confirm the conventional leading order result. 
There have been too many papers since to include as references. 
The only possible problem raised here appears independently of 
these issues and only at fourth order in the neutrino masses.


\begin{thebibliography}{99}
\bibitem{clear} This is a particularly clear example as, aside from the 
branching ratio of almost exactly unity for a charged pion to decay into 
a muon, determination of the recoil particle energy and momentum to 
quite modest accuracy will suffice to determine the invariant mass 
sufficiently well to distinguish a muon from a decay with a recoiling 
electron which only occurs with a branching ratio of $\approx 10^{-4}$ 
anyway.
\end{thebibliography}
\end{document}